\documentclass[traditabstract]{aa} % for the abstract without structuration 
\usepackage{graphicx}
\usepackage{txfonts}
\usepackage{aalongtable}
\begin{document}

\title{An asteroseismic study of the O9V star HD\,46202 from CoRoT space-based
  photometry\thanks{The CoRoT space mission was developed and is operated by the
    French space agency CNES, with the participation of ESA's RSSD and Science
    Programmes, Austria, Belgium, Brazil, Germany, and Spain.}}
%   \subtitle{}

\author{M. Briquet\inst{1,}\thanks{Postdoctoral Fellow of the Fund for
    Scientific Research of Flanders (FWO), Belgium}
          \and
          C. Aerts\inst{1,2}
          \and
          A. Baglin\inst{3}
          \and
          M. F. Nieva\inst{4}
          \and
          P. Degroote\inst{1}
          \and
          N. Przybilla\inst{5}
          \and
          A. Noels\inst{6}
          \and
          F. Schiller\inst{5}
          \and
          M. Vu\v{c}kovi\'c\inst{1}
          \and
          R. Oreiro\inst{1}
          \and      
          K. Smolders\inst{1}
          \and
          M. Auvergne\inst{3}
          \and
          F. Baudin\inst{7}
          \and
          C. Catala\inst{3}
          \and
          E. Michel\inst{3}
          \and
          R. Samadi\inst{3}
          }

\institute{Instituut voor Sterrenkunde, K.U.Leuven, Celestijnenlaan 
200D, B-3001 Leuven, Belgium\\ \email{maryline@ster.kuleuven.be}
\and Department of Astrophysics, IMAPP, University of Nijmegen, PO Box 9010,
6500 GL Nijmegen, The Netherlands
\and LESIA, CNRS UMR8109, Universit\'e Pierre et Marie Curie, Universit\'e Denis
Diderot, Observatoire de Paris, 92195 Meudon Cedex, France
\and
Max Planck Institute for Astrophysics, Karl Schwarzschild Str. 1, Garching bei
M\"unchen D-85741, Germany
\and
Dr.\ Karl Remeis Observatory \& ECAP, University of Erlangen-Nuremberg,
Sternwartstrasse 7, D-96049 Bamberg, Germany
\and Institut d'Astrophysique et de G\'eophysique, University of Li\`ege,
B\^at.\ B5C, All\'ee du 6 Ao\^ut 17, B-4000 Li\`ege, Belgium
\and Institut d'Astrophysique Spatiale (IAS), B\^atiment 121, F-91405, Orsay
Cedex, France }

\date{Received; accepted}

\authorrunning{Briquet et al.}
\titlerunning{An asteroseismic study of the O9V star HD\,46202 from CoRoT space-based
  photometry}

\offprints{maryline@ster.kuleuven.be}

% \abstract{}{}{}{}{} 
% 5 {} token are mandatory
 
  \abstract{
The O9V star HD\,46202, which is a member of the young open cluster NGC
    2244, was observed by the CoRoT satellite in October/November 2008 during a
    short run of 34 days. From the very high-precision light curve, we clearly
    detect $\beta$~Cep-like pulsation frequencies with amplitudes of 
$\sim$0.1~mmag and below. A comparison with stellar models was performed
    using a $\chi^2$ as a measure for the goodness-of-fit between the observed
    and theoretically computed frequencies. The physical parameters of our best-fitting models are compatible with the ones deduced spectroscopically. A core overshooting parameter $\alpha_{ov} = 0.10
    \pm 0.05$ pressure scale height is required. None of the observed
    frequencies are theoretically excited with the input physics used in our
    study. More theoretical work is thus needed to overcome this
    shortcoming in how we understand the excitation mechanism 
of pulsation modes
    in such a massive star. A similar excitation problem has also been encountered for
    certain pulsation modes in $\beta$~Cep stars recently modelled
    asteroseismically.
    }
   \maketitle
%
%________________________________________________________________

\section{Introduction}
Asteroseismology is the domain of astrophysics that probes the internal
structure of stars by using their stellar oscillations. Different classes of
pulsating stars are known across the H-R diagram, including pulsating
main-sequence B-type stars (with masses between 3 and 20 M$_\odot$), such as the
so-called $\beta$~Cep stars and the slowly pulsating B stars (e.g., Aerts et
al.\ \cite{aertsetal}).  Currently, the part of the H-R diagram corresponding to even hotter
and more massive main-sequence stars has been poorly explored for asteroseismic
purposes, both theoretically and observationally.

 Due to lack of firm observational detections, the theory of stellar
  oscillations has hardly been applied to O star models. On the one hand, one
  expects, for late O-type stars such as HD\,46202 (O9V) studied in this
  paper, that oscillation modes excited by the
  $\kappa\,$mechanism occur with periods of several hours, just as for the well-established class of
  $\beta$~Cep stars. The theoretical treatment for the lower mass main-sequence
  O stars is the same as the one for the early-type B stars because stellar
  winds are not expected to play an important role in the oscillatory
  behaviour. Radiation-driven winds 
have thus been ignored when computing instability strips
  (e.g., Pamyatnykh \cite{pamyatnykh}, Miglio et al.\ \cite{miglioab}). On the other hand, theory
  predicts the occurrence of so-called strange mode oscillations in stars close
  to the Humphreys-Davidson limit. These are due to a convection zone in the
  outer stellar layers where a density inversion occurs and the
  radiation pressure dominates (e.g., Glatzel \& Kiriakidis\ \cite{glatzel_kiriakidis}, Papaloizou et al.\ \cite{papaloizou}, Saio et al.\ \cite{saioetal}).  Review papers on the topics are available by
  Glatzel (\cite{glatzel}) and by Saio (\cite{saio}). Observations have not established that there are strange modes in O stars.  

On the observational side, there are a few O-type stars for which spectroscopic
and photometric variability linked to stellar pulsation has been searched for
and found from ground-based data. 

Convincing evidence of stellar pulsation has been detected in $\zeta$~Oph (Kambe et al.\ \cite{kambe}), HD\,93521
(Howarth \& Reid\ \cite{howarth_reid}, Rauw et al.\ \cite{rauw}), $\xi$~Persei, and $\lambda$~Cephei (de Jong et al.\ \cite{dejong}). 
In fact, it is difficult to discover pulsation modes in O-type targets because the observed variability that might come 
from pulsations has very low amplitude (e.g., Henrichs\ \cite{henrichs} for a summary).

In this context clearer detections can be expected from long uninterrupted
photometric time series at micro-magnitude precision obtained from space
satellites such as CoRoT (Convection, Rotation and planetary Transits, Auvergne
et al.\ \cite{auvergne}). As part of the asteroseismology programme, several O-type stars
were observed by CoRoT during the second short run SRa02 pointing towards the
anticentre of the Galaxy, e.g., the O-star binaries HD\,46149 (Degroote et al.\
\cite{degroote10}) and HD\,47129 (Mahy et al.\ \cite{mahy}).  In this paper, we present a study of
the single O9V star HD\,46202, which is a member of the young open cluster
NGC~2244 in the Rosette Nebula.

%__________________________________________________________________

\section{Frequency analysis on CoRoT photometry}

\begin{figure}
\centering
\includegraphics[angle=270,totalheight=0.35\textwidth]{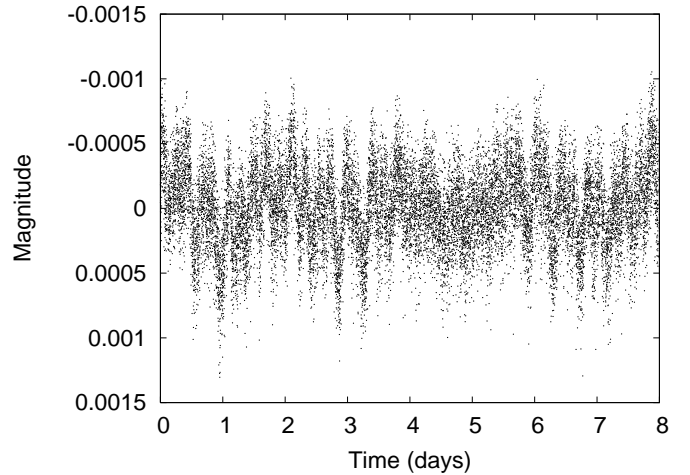}
\caption{Part of the detrended CoRoT light curve, converted to magnitude (mag = -2.5 $\log_{10}$(flux), once the average has been substracted). The first datapoint corresponds to 2454759 HJD.}
\label{lc}
\end{figure}

The CoRoT data of HD\,46202 were taken during 34 days between 2008 October 8 and 2008 November 12 with an average time sampling of 32 seconds. This
involves a frequency resolution of $\sim$0.03\,d$^{-1}$. After removing all
flagged measurements, such as the ones suffering from hot pixels during the
passage through the South Atlantic Anomaly, the light curve consists of 81\,781
datapoints. As in most CoRoT light curves, a downward trend considered to be of
instrumental origin is visible. Moreover, the light curve presents three jumps
during the last five days of observation. We divided the light curve into four parts
and, for each of these selected regions, we removed a trend using a
linear approximation. To visualise the variability, only part of the detrended
light curve is shown in Fig.\,\ref{lc}. We can clearly see a variation with a
period of the order of 2 days. Moreover, variations on a shorter time scale are
also observed.

\begin{figure}
\centering
\includegraphics[angle=270,totalheight=0.35\textwidth]{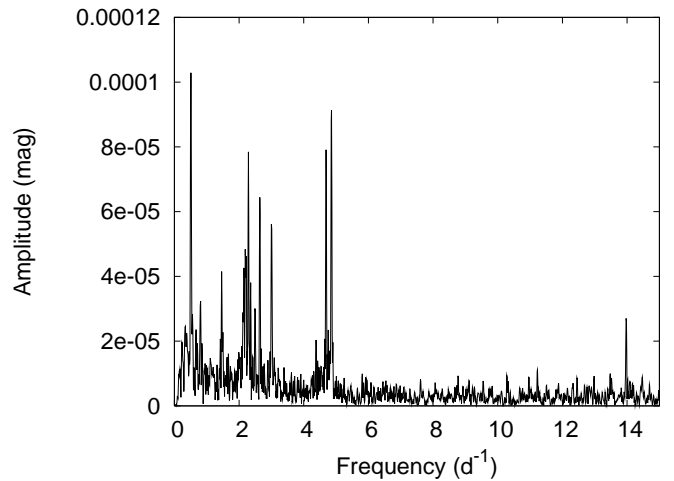}
\caption{Scargle periodogram of the detrended CoRoT light curve, converted to magnitude.}
\label{fourier}
\end{figure}

The Scargle periodogram (Scargle\ \cite{scargle}) of the detrended light curve is presented
in Fig.\,\ref{fourier}. Clear peaks are detected between 0.5 and
5\,d$^{-1}$. The peak at 13.97\,d$^{-1}$ is not intrinsic to the star but
corresponds to the orbital frequency of the CoRoT satellite. To determine the
whole frequency spectrum of the star we used the methodology described in
Degroote et al. (\cite{degroote09a}), who analysed numerous B-type pulsators using light
curves from the exoplanet data of the CoRoT satellite. This methodology uses the
usual method of successive prewhitening, along with a statistical stop
criterion based on the one first introduced by Scargle (\cite{scargle}). The statistical
stop criterion defined in Degroote et al.\ (\cite{degroote09a}) was adopted 
instead of a signal-to-noise (S/N) criterion (Breger et
al.\ \cite{breger}). The reason is that, although a signal originating from 
nonsinusoidal processes (e.g., red noise) is also considered to be significant,
and thus prewhitened using a sum-of-sines model, isolated peaks at lower
amplitudes can be recovered outside of these regions.

The frequencies with the highest amplitudes, which turn out to be typical 
$\beta$\,Cep-like pulsation modes, are listed in Table\,\ref{freq}. In addition,
all the
frequencies with an amplitude lower than the one corresponding to the
satellite's orbital frequency are given in the Appendix. In total 67 frequencies
are found to be significant by our adopted stop criterion. For comparison
purposes, we also indicate the signal-to-noise (S/N) ratio, which is computed
using a 6\,d$^{-1}$ box-interval centred on the considered frequency. The error estimate on the frequency is computed as in Degroote et al.\ (\cite{degroote09a}), i.e., using the formula of Montgomery \& O'Donoghue (\cite{montgomery_odonoghue}), but also taking correlation effects into account according to the formalism by Schwarzenberg-Czerny (\cite{schwarzenberg-czerny}).

\begin{table}
\caption{The highest-amplitude frequencies derived from the detrended CoRoT light curve after subsequent prewhitening, their corresponding amplitude, and signal-to-noise ratio.}
\begin{center} 
\begin{tabular}{ccccc}
\hline\hline
ID  & $f$(d$^{-1})$ & $f$($\mu$Hz) & Amplitude (mmag) & S/N ratio \\
\hline
$f_1$ & 0.510 &  5.901& 0.1079 &  9.3\\
$f_2$ & 4.856  & 56.204& 0.0934 &  11.7\\
$f_3$ & 4.691 & 54.295 & 0.0798 &  10.7\\
$f_4$ & 2.290 & 26.500 & 0.0708 & 8.5 \\
$f_5$ & 2.643 & 30.586 & 0.0645 &  7.7\\
$f_6$ & 3.004 & 34.769 & 0.0587 &  7.2\\
$f_7$ & 2.195 & 25.404 & 0.0460 &  5.7\\
$f_8$ & 1.462 & 16.919 & 0.0426  &  5.6\\
$f_9$ & 2.222 & 25.722 & 0.0339 &  4.8\\
$f_{10}$ & 0.804 & 9.305 & 0.0314  & 4.5 \\
$f_{11}$ & 2.127 & 24.613 & 0.0307 & 4.3 \\
$f_{12}^{*}$ & 13.973 & 161.728 & 0.0298 & 10.6 \\
\hline
\end{tabular}
\end{center}
\begin{list}{}{}
\item[$^{*}$]the satellite's orbital frequency 
\end{list}
\label{freq}

\end{table}

\begin{figure}
\centering
\includegraphics[angle=0,totalheight=0.35\textwidth]{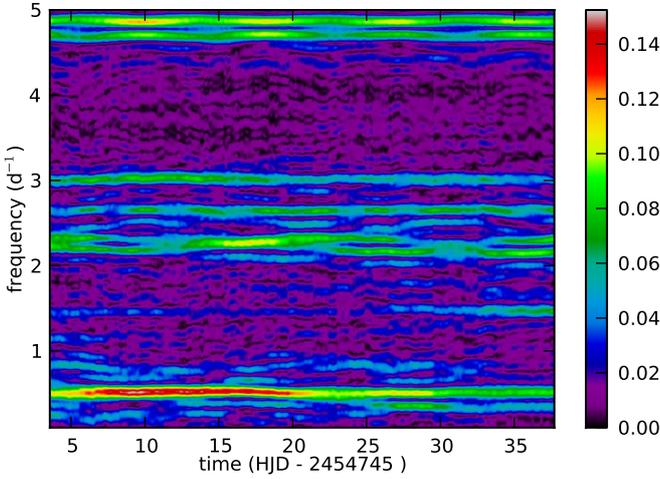}
\caption{The short time Fourier transform of the detrended CoRoT light curve, computed with a window width of 10 days. The colour scale is in mmag units.}
\label{stft}
\end{figure}

Recent studies of CoRoT light curves of B-type stars highlighted variations in
the amplitude of frequencies and in the frequencies themselves as a function of
time (e.g., in the $\beta$\,Cep star HD\,180642, Degroote et al.\ \cite{degroote09b}; 
in the B0.5IVe star HD\,49330, Huat et al.\ \cite{huat}).
In order to check the stability of the
frequencies of HD\,46202 listed in Table\,\ref{freq}, we applied a short time
Fourier transform. Different window widths
were used, leading to the conclusion that the frequencies are stable in
time. There is no convincing proof of changing amplitudes either. The detected
low amplitude variations in time are typical of beating effects for a
multiperiodic pulsator (Fig.\,\ref{stft}).

We searched for rotational frequency splittings with the automatic methods
described in Degroote et al.\ (\cite{degroote09a}), but we could not unambiguously or
confidently find such multiplets. For the highest amplitude modes, the
frequencies $f_{4}$, $f_{5}$, and $f_{6}$ (see Table\,\ref{freq}) might belong to
the same multiplet and the frequencies $f_{8}$, $f_{10}$, and $f_{11}$ might
belong to another one.
However, recent work on the $\beta$~Cep star 12~Lacertae
taught us to be cautious in attributing frequencies to multiplets
without empirical identification of the mode degree. For this
latter target, Dziembowski \&
Jerzykiewicz (\cite{dziembowski_jerzykiewicz}) assumed 
that three of the strongest modes, almost equidistant, belong
to the same multiplet in their seismic modelling. 
Later, a photometric mode identification by Handler et
al.\ (\cite{handler}) ruled out this assumption by proving that the three modes are
actually associated with three different values of the degree $\ell$. It is
obvious that reliable empirical mode identification from multicolour photometry,
and/or spectroscopy, is indispensable for definite conclusions.

CoRoT observations also revealed solar-like
oscillations in two massive stars: the $\beta$~Cep star HD\,180642 (Belkacem et
al.\ \cite{belkacem}) and the O-type binary HD 46149 (Degroote et al.\ \cite{degroote10}). To
search for the signature of such oscillations in HD\,46202, we
performed an autocorrelation of the time series and periodogram, both for the
original detrended light curve and for the light curve prewhitened with the
highest-amplitude frequencies. We could not find any structure typical of
solar-like oscillations with amplitudes above 0.009 mmag.

\section{Search for variability in spectra}
\begin{figure}
\centering
\includegraphics[angle=270,totalheight=0.35\textwidth]{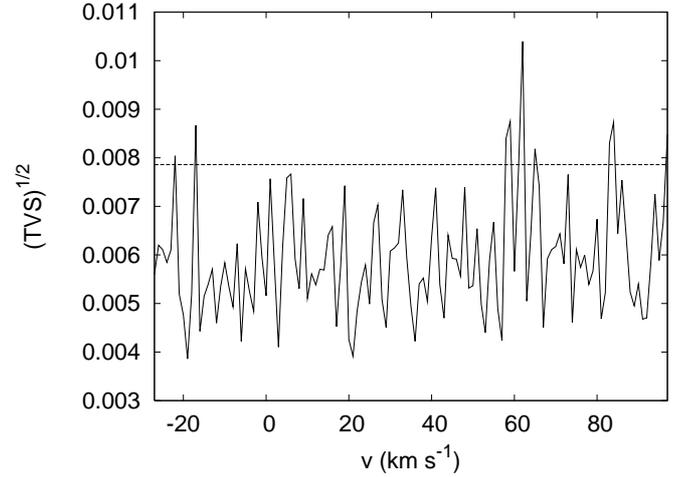}
\caption{Temporal variance spectrum (TVS) computed from the cross-correlated profiles computed by taking all elements into account, except H. The line corresponds to the 99\% significance level for the variability evaluated following the approach of Fullerton et al.\,(\cite{fullerton}).}
\label{tvs}
\end{figure} 

From the photometric amplitudes of the pulsation modes being 0.1 mmag or less, it is
clear that these modes cannot be detectable in spectra of moderate
S/N. Nevertheless, the study of a time series of spectral lines
remains relevant when searching for other kinds of variability. First, we
can check for the presence of a variable stellar wind. Also, modes of very high-degree $\ell$ might be visible in the spectral lines (e.g., as in the
$\beta\,$Cep star $\omega^1\,$Sco, Telting \& Schrijvers\ \cite{telting_schrijvers}), although they are
not observed in the CoRoT light curve. Additionally, some spectral lines
associated to only certain elements might be variable with a periodicity linked
to the stellar rotation and magnetic field.

Consequently, we gathered 44 high-resolution spectra (R = 50\,000, S/N of $\sim$60) with the CORALIE \'echelle spectrograph
attached to the 1.2m Leonard Euler telescope (La Silla, Chile). First,
variability was searched in individual line profiles. Then, to increase the S/N
of the spectra and improve the detection of the line-profile variations, we used
cross-correlated profiles computed using the least squares deconvolution (LSD)
method (Donati et al.\ \cite{donati}) and taking all elements into account except H. We also computed cross-correlated profiles associated with only one element for
all possible elements, such as helium, silicon, nitrogen and oxygen. To
perform our frequency analysis, we used the software package
FAMIAS\footnote{FAMIAS has been developed in the framework of the FP6 European
  Coordination Action HELAS – http://www.helas-eu.org/} (Frequency Analysis and
Mode Identification for Asteroseismology, Zima\ \cite{zima}). 
The significance of variability in stellar spectra can be quantified by a temporal variance spectrum (TVS, Fullerton et al.\,\cite{fullerton}). We computed TVS diagrams for individual line profiles and cross-correlated profiles (Rauw et al.\,\cite{rauw01}). One of them is illustrated in Fig.\,\ref{tvs}. Even with such state-of-the-art techniques, no spectroscopic variability could be detected. 

\begin{figure*}
\sidecaption
\includegraphics[width=12cm]{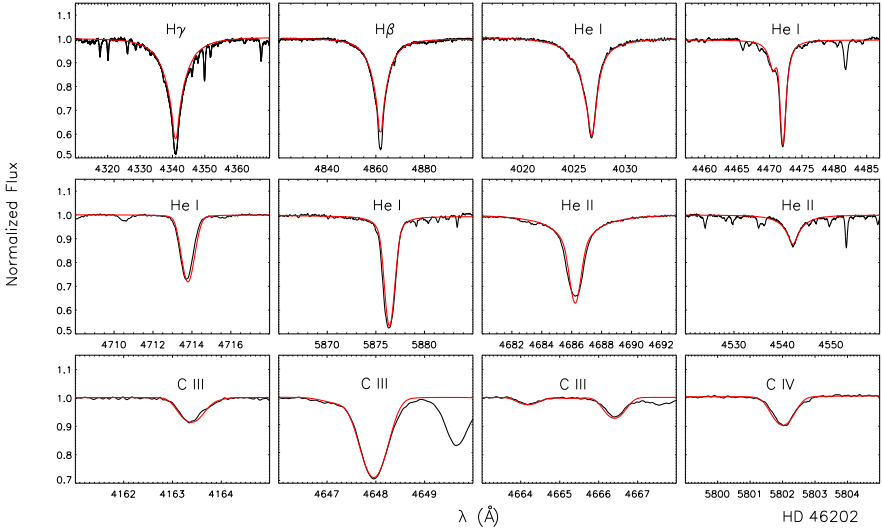}
\caption{Examples of line fits (red) for H, He, and C to the high-quality observed spectrum (black). 
The hydrogen and helium lines are reproduced with a single
synthetic spectrum, while for carbon a slightly different abundance value
is adopted per line, giving rise to a 1-$\sigma$ uncertainty of 0.10\,dex
due to line-to-line scatter.}
\label{fit}
\end{figure*} 

\section{Fundamental parameters}
In the next section, we compute stellar models for our target, which match 
the pulsation frequencies discovered in the CoRoT photometry. The fundamental parameters and 
chemical composition of HD\,46202 are also used as additional constraints. To deduce the 
latter, we used a high-quality spectrum with extended wavelength coverage, taken with the 
ESO/MPG 2.2m telescope with FEROS (Kaufer et al.~\cite{Kaufer99}). Six individual spectra, 
observed between 2008 April 16--26 of 720\,s exposure time each were extracted from the 
ESO archive and reduced with the MIDAS pipeline. Coaddition of these data gave a spectrum 
with $S/N$ of more than 500 per resolution element in the optical blue region.

\begin{table*}
\caption{Atmospheric parameters, photometric properties and
fundamental parameters of HD\,46202 as derived from the spectroscopic
analysis.}
\begin{center}
\begin{tabular}{lr@{\hspace{1.2cm}}lr@{\hspace{1.2cm}}lr}
\hline\hline
$T_\mathrm{eff}$\,(K)             & 34100$\pm$600 & $V$              & 8.18$\pm$0.02    & $M$/$M_\odot$               & 19.6$\pm$2.7\\
$\log g$\,(cgs)                   & 4.17$\pm$0.07 & $B-V$            & 0.17$\pm$0.02    & $R$/$R_\odot$               & 6.5$^{+0.9}_{-0.7}$\\
$\xi$\,(km\,s$^{-1}$)             & 6$\pm$2       & $E(B-V)$         & 0.49$\pm$0.03    & $\log L/L_\odot$            & 4.71$\pm$0.10\\
$v\sin i$\,(km\,s$^{-1}$)         & 25$\pm$7      & $R_V$            & 2.7$\pm$0.1      & $X$\,(by mass)              & 0.715$\pm$0.033\\
$\zeta$\,(km\,s$^{-1}$)           & 15$\pm$7      & $M_V$            & $-$3.73$\pm$0.25 & $Y$\,(by mass)              & 0.271$\pm$0.033\\
$\dot{M}$\,($M_\odot$\,yr$^{-1}$) & $<$10$^{-7}$  & $B.C.$           & $-$3.30$\pm$0.05 & $Z$\,(by mass)              & 0.014$\pm$0.001\\
$d_\mathrm{spec}$\,(kpc)          & 1.3$\pm$0.2   & $M_\mathrm{bol}$ & $-$7.03$\pm$0.26 & $\tau_\mathrm{evol}$\,(Myr) & 2$\pm$1.5\\[1mm]
\hline
\end{tabular}
\end{center}
\label{parameters}
\end{table*}

Our quantitative spectral analysis follows the methodology described by Nieva \& 
Przybilla~(\cite{NP07}, \cite{NP08}). In brief, non-LTE line-formation calculations were 
performed with the codes {\sc Detail} and {\sc Surface} (Giddings~\cite{gid81}; 
Butler \& Giddings~\cite{but_gid85}; both updated by K. Butler) on the basis of a 
prescribed LTE atmospheric structure ({\sc Atlas9}, Kurucz~\cite{kur93}) and accounting for 
the most recent model atoms of Przybilla et al.~(\cite{P08}). The
method is equivalent to full non-LTE model atmosphere analyses for
late O-type stars with weak winds (Nieva \& Przybilla~\cite{NP07}).

The parameters were determined via simultaneous fitting of all observed H lines 
and ionisation equilibria of \ion{He}{i/ii} and \ion{C}{iii/iv}. Our
results from the spectral fitting are summarised in Table~\ref{parameters} for the 
effective temperature $T_\mathrm{eff}$, surface gravity  $\log g$, microturbulence $\xi$, 
and the projected rotational $v\sin i$ and macroturbulent velocity $\zeta$.
Examples of our fits to the observed spectrum are shown in 
Fig.~\ref{fit} for these final parameters. Additional models were calculated with the 
non-LTE code {\sc Fastwind} (Puls et al.~\cite{Puls05}), which accounts for non-LTE metal 
line blanketing, mass loss, and spherical geometry, for an independent test of 
the parameter determination. Similar atmospheric parameters were derived from observed 
H and \ion{He}{i/ii} line spectra\footnote{A comprehensive analysis of the metal line 
spectrum is not feasible with {\sc Fastwind} at present.}, 
confirming the match of our hydrostatic approach also with hydrodynamical models for late 
O-type main sequence stars with weak stellar winds. The {\sc Fastwind}
calculations also allowed the mass-loss rate $\dot{M}$ from H$\alpha$ line-profile fits 
to be constrained. The upper limit found can be considered as insignificant in our comparison 
with stellar evolutionary models.

The helium, carbon, nitrogen, and magnesium abundances derived with 
{\sc Detail}/{\sc Surface} coincide with 
the cosmic abundance standard (CAS) from early B-type stars in the solar neighbourhood 
as proposed by Przybilla et al.~(\cite{P08}). From an extension of that work (Nieva \& 
Przybilla, in preparation) we found evidence that abundance analyses of O, Si, and Fe 
for late O-type stars require extensions of the model atoms, which is beyond the scope 
of the present work. We therefore adopt CAS values for these elements,
implying mass fractions for hydrogen $X$, helium $Y$, and the metals
$Z$ in HD\,46202 as indicated in Table~\ref{parameters}, impli\-citly
assuming that the present surface abundances are representative 
for the pristine values.

\begin{figure}
\centering
\includegraphics[width=.99\linewidth]{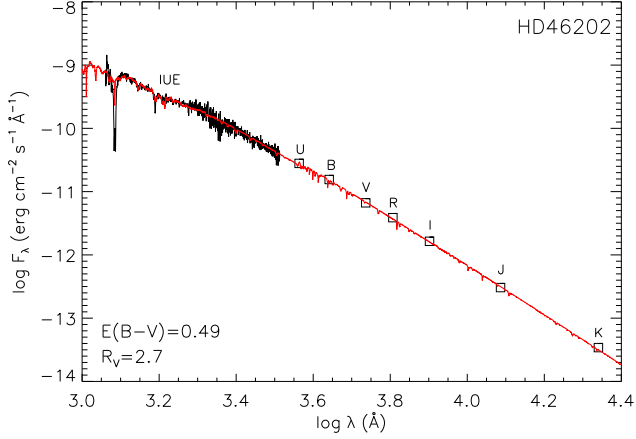}
\caption{Fit of the {\sc Detail} non-LTE model flux for our finally 
adopted parameters from Table~\ref{parameters} (red
line) to the observed spectral energy distribution of HD\,46202 
(black line, squares) from the UV to the near-IR. The observations have 
been dereddened and normalised in $V$.}
\label{SED}
\end{figure}

To verify the results from our spectroscopic analysis, we
compared our non-LTE model flux with the observed spectral energy
distribution of HD\,46202 in Fig.~\ref{SED}. For this UV
spectrophotometry was extracted from the IUE archive, and Johnson
magnitudes were adopted from Morel \& Magnenat~(\cite{Morel78}). The
latter were transformed into absolute fluxes using zeropoints from Bessel 
et al.~(\cite{Bessel98}). The observed flux was dereddened according to 
the reddening law described by Cardelli et al.~(\cite{Cardelli89}),
with both the colour excess $E(B-V)$ and the ratio of total-to-selective
extinction $R_V$ requiring determination. The data were
normalised in $V$. An excellent match of theory with observation is
obtained for the values given in Table~\ref{parameters}. 
The strong Ly$\alpha$ line mostly has interstellar origins.

This allowed the dereddened apparent visual magnitude to be determined
and the spectroscopic distance $d_\mathrm{spec}$ to be derived
following Ramspeck et al.~(\cite{Ramspeck01}). The required stellar
mass $M$ was obtained by comparison of the star's position in the
$\log T_\mathrm{eff}$--$\log g$ diagram with evolutionary tracks computed as described in Sect.\,5 (also the
evolutionary age $\tau_\mathrm{evol}$ was estimated this way). 
This in turn facilitated the absolute visual magnitude $M_V$ to be determined,
along with the bolometric magnitude $M_\mathrm{bol}$, the stellar luminosity $L$, and the stellar radius $R$, by using the bolometric correction $B.C.$ from our model. Thus, a fully consistent
{\em spectroscopic solution} for the fundamental stellar parameters of HD\,46202 was
established, as summarised in Table~\ref{parameters}. 

In consequence, HD\,46202 is an unevolved star close to the 
zero-age main sequence, which shows no signature of mixing with
CN-processed material (Przybilla et al.~\cite{P10}).
The derived colour excess, $R_V$, distance, and age are in excellent
agreement with the NGC\,2244 cluster properties determined by Hensberge et
al.~(\cite{Hensberge00}).

\section{Comparison with stellar models}

We checked that state-of-the-art stellar models with standard
physics can account for the frequency spectrum observed by CoRoT. When $X$ and
$Z$ are fixed, we attempted to asteroseismically constrain the model parameters of HD\,46202,
which are the mass, the central hydrogen abundance (or age), and the core
convective overshooting parameter.

\subsection{Numerical tools and model input physics}\label{physics}

Stellar models for non-rotating stars were computed with the evolutionary code
CL\'ES (Code Li\'egeois d'\'Evolution Stellaire, Scuflaire et al.\
\cite{scuflaire}). We used the OPAL2001 equation of state (Rogers \& Nayfonov\
\cite{rogers}; Caughlan \& Fowler\ \cite{caughlan}), with nuclear reaction rates
from Formicola et al.\ (\cite{formicola}) for the $^{14}$N$(p,\gamma)^{15}$O
cross-section. Convective transport is treated by using the classical mixing
length theory of convection (B\"ohm-Vitense\ \cite{bohm}). For the chemical
composition, we used the solar mixture from Asplund et al.\ (\cite{asplund}). We
used OP opacity tables (Seaton \ \cite{seaton}) computed for this mixture. These tables are completed at $\log \rm{T} < 4.1$ with the low-temperature tables of Ferguson et al.\ (\cite{ferguson}).

When $X \in [0.682,0.748]$ and $Z \in [0.013,0.015]$ were adopted (Przybilla et al.~\cite{P08}), a fine grid of stellar evolutionary
models was computed for a mass between 16 and 28 M$_\odot$ in steps of 0.1
M$_\odot$ and a core convective overshooting parameter $\alpha_{\rm ov}$ between 0 and 0.5
pressure scale heights in steps of 0.05. For each main-sequence stellar model,
we calculated the theoretical frequency spectrum of low-order p- and g-modes
with a degree of the oscillation up to $\ell = 4$ using a standard adiabatic
code for non-rotating stellar models (Scuflaire et al.\ \cite{scuflaire2}).

\subsection{Matching the highest-amplitude frequencies}

\begin{figure}
\centering
\includegraphics[angle=270,totalheight=0.35\textwidth]{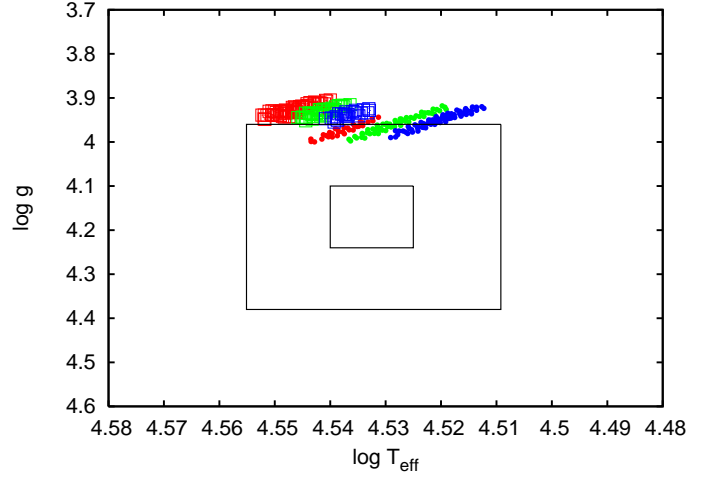}
\caption{The 1-$\sigma$ and 3-$\sigma$ error boxes represent the position of HD\,46202, deduced from spectroscopy, in the $\log T_\mathrm{eff}$--$\log g$ diagram. The positions of all the asteroseismic models having a $\chi^2 \le 2$ are also shown, for $X=0.682$ (in red), $X=0.715$ (in green), and $X=0.748$ (in blue). We refer to the text for more explanations.}
\label{chi2}
\end{figure} 

To each model of our computed grid of stellar evolutionary models, we assigned a
$\chi^2$-value, which compares the observed frequencies $f_i^{\rm obs}$ to the theoretically
computed ones $f_i^{\rm th}$, as follows: 
\begin{equation}
\centering
\chi^2 = \sum_{i=1}^{N}\frac{(f_i^{\rm obs}-f_i^{\rm th})^2}{(\sigma_i^{\rm obs})^2}
\end{equation}
where $\sigma_i^{\rm obs}$ is the observational uncertainty.
The lower the $\chi^2$-value, the better the match between
observed and predicted frequencies. We selected the observed frequencies with
only the highest amplitudes, as listed in Table\,\ref{freq}. The
$\chi^2$-value is computed, and the theoretical frequency value is the
closest to the observed one. Because of the lack of mode identification, no
restriction on the degree $\ell$ was imposed on the theoretical frequencies,
which correspond to zonal modes ($m = 0$).

In Fig.\,\ref{chi2} the positions of all the models having a $\chi^2 \le 2$,
without any a priori restriction on $\log T_{\rm eff}$, but with $\log g \ge 3.89$, 
are compared to the error box representing the
position of HD\,46202 in the $\log T_\mathrm{eff}$--$\log g$ diagram, as spectroscopically deduced.

We conclude that the positions of our best-fitting models are compatible with the spectroscopic one. Indeed, the discrepancy found in log g can be considered as small in view of current uncertainties in the theory of stellar structure and stellar atmospheres. Despite great efforts have been made to identify and reduce numerous systematic errors of atmospheric parameters related to the data reduction, spectral modelling, and spectral analysis (see Nieva \& Przybilla\,\cite{np10} for a summary), the remaining systematics cannot be excluded.

One of the main factors in this context is the unknown true rotational velocity $v_\mathrm{rot}$ of HD\,46202. The star shows an exceptionally low $v \sin i$ for the massive star population of NGC\,2244, where about 30\% of the stars have $v \sin i$\,$>$\,250\,km\,s$^{-1}$ (Huang \& Gies\,\cite{huang_gies}), indicating a large fraction of truly fast rotators. The measured $v \sin i$ of HD\,46202 is only a lower limit, since $v_\mathrm{rot}$ may reach much higher values if the star is seen nearly pole-on. The spectroscopically derived $\log g$ would be the polar value in that case, while the average $\log g$ on the surface of the rotationally deformed star could be significantly lower. Fast rotation would require not only a re-interpretation of the spectroscopically derived parameters but also consideration of non-standard physics in the stellar structure and evolutionary models (see e.g. Ekstr\"om et al.\,\cite{ekstrom}), with additional consequences for the predicted frequency spectrum and its excitation. We conclude that the potential nature of HD\,46202 as a fast rotator seen nearly pole-on could resolve the remaining discrepancy between the spectroscopic and asteroseismic solutions. The way to check this would be to gather a very high-resolution time series (typically R$\sim$100\,000 of high S/N ($>$ 300) spectroscopy covering the beat pattern of the oscillations). In this way, a spectroscopic mode identification can be achieved with certainty and, along with it, an estimate of the inclination angle of the star. 
  
We note that a similar shift in $\log g$ was also obtained for the $\beta$~Cep star $\theta$~Ophiuchi (Briquet et al.\ \cite{briquet}). For this target, the identification of a fundamental radial mode could pinpoint the seismic $\log g$-value while the spectroscopic $\log g$-value was found to be about 0.15 dex higher.  

\subsection{Most probable model parameters}

In Fig.\,\ref{chi2}, we distinguish two groups of models, shown as open squares and as filled circles. 
Looking deeper into the theoretical frequency spectrum of these two groups, we conclude that the most probable models are those represented as open squares in Fig.\,\ref{chi2}. Our arguments are the following.

For the models shown as filled circles, the theoretical frequencies matching the
observed frequencies of Table\,\ref{freq} correspond to $\ell = 3$ or $\ell = 4$
modes only, except for the frequency $f_6$ which has an $\ell = 1$. It is very
unlikely that no mode with $\ell=0$ and $\ell = 2$ is observed among the highest amplitudes modes. The reverse scenario, i.e. with only $\ell = 0, 1, 2$ modes among
the highest amplitude modes, is certainly more plausible for reasons of
cancellation effects for higher degree modes.

The models shown as open squares fit $f_2 = 4.856$\,d$^{-1}$ as the fundamental
radial mode, $f_3 = 4.691$\,d$^{-1}$ as an $(\ell,n) = (2,-1)$ mode and $f_6 =
3.004$\,d$^{-1}$ as an $(\ell,n) = (1,-1)$ mode. For the modes with lower frequency
values, the theoretical frequency spectrum is denser, so an unambiguous
identification is more difficult. Our best-fitting model has a mass between 23.3 and 24.9
M$_{\odot}$ and a core convective overshooting parameter between 0.05 and 0.15 pressure scale heights. The other physical parameters are given in Table\,\ref{model_par}.

\begin{table}
\caption{Physical parameters of HD\,46202 as derived from the asteroseismic modelling, $X \in [0.682,0.748]$ and $Z \in [0.013,0.015]$. }
\begin{center}
\begin{tabular}{cc}
\hline\hline
$T_\mathrm{eff}$\,(K)  &   34800$\pm$780        \\
$\log g$\,(cgs)       &   3.93$\pm$0.03         \\
$M$/$M_\odot$         &   24.1$\pm$0.8          \\
$\tau_\mathrm{evol}$\,(Myr) & 4.3$\pm$0.5      \\
$\alpha_{\rm ov}$     &   0.10$\pm$0.05          \\[1mm]
\hline
\end{tabular}
\end{center}
\label{model_par}
\end{table}

\subsection{Checking the excitation of the modes}

For all models represented in Fig.\,\ref{chi2}, we checked the
excitation of the pulsation modes with the linear non-adiabatic code MAD
developed by Dupret et al.\,(\cite{dupret}). We found that none of the observed
frequencies is theoretically excited using the input physics described in
Sect.\,\ref{physics}.

Such an excitation problem has already been encountered for other massive stars. For
instance, the two $\beta$\,Cep stars 12~Lacertae (B2III) and 
$\nu$~Eridani (B2III) present 
observed modes that have not yet been predicted by pulsation models, although
different attempts to solve the discrepancies between theory and observations
have been suggested in the literature. We refer to Dziembowski \& Pamyatnykh
(\cite{dziembowski_pamyatnykh}), and references therein, for the latest discussion of the matter. Another
choice of opacity table (e.g. OPAL opacities) and/or metal mixture
(e.g. Grevesse \& Noels\ \cite{grevesse_noels}) does not reconcile the observed and excited model
frequencies for HD\,46202. Miglio et al.\ (\cite{miglioab}) show that more modes are
found to be excited in pulsating B-type stars when using OP opacities with the
Asplund et al.\ (\cite{asplund}) mixture, as adopted in our study, compared to the case
where the OPAL opacities are considered. To solve shortcomings
in our understanding of excitation mechanisms in massive stars and, in HD\,46202
in particular, is beyond the scope of this paper but deserves to be addressed in depth. One explanation might be that current opacities are still
underestimated in the region where the driving of pulsation modes occurs.

\section{Conclusions} 

The space white-light CoRoT photometry reveals the presence of $\beta$~Cep-like
pulsations of very small amplitudes ($\sim$0.1 mmag and below) in the O9V star
HD\,46202. By means of CoRoT data, variability associated to stellar pulsations,
albeit of a different nature, was also detected in the O-type binary HD\,46149
(Degroote et al.\ \cite{degroote10}) and in Plaskett's star (Mahy et al.\ \cite{mahy}). These
discoveries unambiguously prove there are stellar oscillations in O-type
stars, while opening the way for promising asteroseismic modelling of this kind of
massive star.

For HD\,46202, we found stellar models compatible with the observed
asteroseismic (pulsation frequencies) and non-asteroseismic constraints
(effective temperature and surface gravity). For the model that
best explains the observed oscillations, the frequency $f_2=4.856$\,d$^{-1}$ is
identified as the fundamental radial mode. Empirical mode identification is
desirable in order to confirm this conclusion and to identify the
wavenumbers $(\ell,m)$ of other modes. This task is, however, challenging
because of the very low amplitude of the modes.

Another challenge concerns the excitation of the modes. Even using OP opacities
with the Asplund et al.\ (\cite{asplund}) solar mixture, {\it none\/} of the observed
frequencies of HD\,46202 is predicted to be excited by non-adiabatic
computations, which explain, at least in general, the excitation of the
modes in the less massive pulsating B stars.  Miglio et al.\ (\cite{miglioab})
extensively explored the impact of the choice of opacity tables and metal
mixtures on the instability domains for stars with masses between 2.5 and 18
M$_\odot$ and for different metallicity values $Z$. Similar studies for higher
masses, also including mass loss, are necessary for a comparison with the newly
discovered oscillations in massive O-type stars from space data with
unprecedented quality. Our observational results presented here form a good
  basis for such a theoretical study. Moreover, a detailed exploration of the
  physics in the outer layers leading to the opacity-driven modes detected in
  the O9V star HD\,46202 and the modes of stochastic nature detected in the O8.5V
  star HD\,46149 will allow an in-depth evaluation of the various excitation
  mechanisms that seem to be at work in slowly rotating O-type stars in the
  main-sequence phase.

\begin{acknowledgements}
  We thank Andrea Miglio for providing us his code to compute the $\chi^2$
  function. The research leading to these results has received funding from the
  European Research Council under the European Community's Seventh Framework
  Programme (FP7/2007--2013)/ERC grant agreement n$^\circ$227224 (PROSPERITY),
  from the Research Council of K.U.Leuven, from the Fund for
  Scientific Research of Flanders, and from the Belgian federal
  science policy office Belspo. F.\,S. acknowledges financial support by the ``Studienstiftung des deutschen Volkes''. We thank the GAPHE team of the department of Astrophysics, Geophysics and Oceanography of the University of Li\`ege for providing us their code to compute the temporal variance spectrum.

\end{acknowledgements}

\appendix
\section{The frequencies of HD\,46202 detected in the CoRoT light curve}
\begin{longtable}{ccccccc}
\caption{
    The frequencies derived from the detrended CoRoT light curve after
    subsequent prewhitening, their corresponding error, amplitude, phase, and
    signal-to-noise ratio. The signal is written as $\sum_{i=1}^{67} A_i \sin[2\pi (f_i (t-t_0) + \phi_i)]$-0.00000169, where $t_0 = 2451545$ HJD.
  }\\
\hline\hline
ID  & $f$(d$^{-1})$ & Error & $f$($\mu$Hz) & A(mag) & $\phi (2\pi$\ rad)& S/N \\
\hline
\endfirsthead
\caption{continued.}\\
\hline\hline
ID  & $f$(d$^{-1})$ & Error & $f$($\mu$Hz) & A(mag) & $\phi (2\pi$\ rad)& S/N \\
\hline
\endhead
\hline
$f_{1}$ &  0.50982 & 0.00023 &    5.90069 & 0.000108 &  0.174 &  9.28   \\
$f_{2}$ &  4.85606 & 0.00025 &   56.20440 & 0.000093 &  0.087 & 11.68   \\
$f_{3}$ &  4.69107 & 0.00029 &   54.29479 & 0.000080 & -0.292 & 10.65   \\
$f_{4}$ &  2.28961 & 0.00029 &   26.50012 & 0.000071 &  0.048 &  8.54   \\
$f_{5}$ &  2.64259 & 0.00033 &   30.58553 & 0.000064 &  0.369 &  7.69   \\
$f_{6}$ &  3.00402 & 0.00037 &   34.76875 & 0.000059 & -0.421 &  7.23   \\
$f_{7}$ &  2.19493 & 0.00048 &   25.40428 & 0.000046 & -0.372 &  5.72   \\
$f_{8}$ &  1.46182 & 0.00051 &   16.91921 & 0.000043 &  0.265 &  5.59   \\
$f_{9}$ &  2.22240 & 0.00060 &   25.72222 & 0.000034 &  0.446 &  4.84   \\
$f_{10}$ &  0.80396 & 0.00067 &    9.30509 & 0.000031 & -0.353 &  4.46   \\
$f_{11}$ &  2.12660 & 0.00071 &   24.61343 & 0.000031 &  0.256 &  4.30   \\
$f_{12}$ & 13.97330 & 0.00076 &  161.72801 & 0.000030 & -0.121 & 10.60   \\
$f_{13}$ &  2.15488 & 0.00078 &   24.94074 & 0.000028 & -0.072 &  3.95   \\
$f_{14}$ &  2.35706 & 0.00077 &   27.28079 & 0.000028 & -0.205 &  4.09   \\
$f_{15}$ &  2.49231 & 0.00079 &   28.84618 & 0.000027 & -0.115 &  4.09   \\
$f_{16}$ &  0.54203 & 0.00082 &    6.27350 & 0.000026 & -0.389 &  4.00   \\
$f_{17}$ &  0.57365 & 0.00084 &    6.63947 & 0.000022 & -0.016 &  4.03   \\
$f_{18}$ &  0.32639 & 0.00086 &    3.77766 & 0.000031 &  0.392 &  3.98   \\
$f_{19}$ &  0.40030 & 0.00096 &    4.63310 & 0.000027 & -0.016 &  3.61   \\
$f_{20}$ &  0.22654 & 0.00100 &    2.62199 & 0.000021 &  0.148 &  3.52   \\
$f_{21}$ &  0.35462 & 0.00101 &    4.10440 & 0.000023 &  0.176 &  3.59   \\
$f_{22}$ &  0.66703 & 0.00106 &    7.72025 & 0.000019 & -0.434 &  3.46   \\
$f_{23}$ &  3.08004 & 0.00110 &   35.64861 & 0.000018 &  0.168 &  3.39   \\
$f_{24}$ &  0.85580 & 0.00115 &    9.90509 & 0.000016 &  0.409 &  3.31   \\
$f_{25}$ &  4.38227 & 0.00128 &   50.72072 & 0.000016 &  0.199 &  3.02   \\
$f_{26}$ &  1.48753 & 0.00128 &   17.21678 & 0.000017 &  0.077 &  3.06   \\
$f_{27}$ &  2.00846 & 0.00129 &   23.24606 & 0.000018 & -0.424 &  3.07   \\
$f_{28}$ &  0.99298 & 0.00130 &   11.49282 & 0.000017 &  0.275 &  3.07   \\
$f_{29}$ &  1.11162 & 0.00141 &   12.86597 & 0.000015 &  0.072 &  2.90   \\
$f_{30}$ &  0.71597 & 0.00145 &    8.28669 & 0.000015 &  0.496 &  2.86   \\
$f_{31}$ &  2.07063 & 0.00145 &   23.96562 & 0.000015 &  0.252 &  2.89   \\
$f_{32}$ &  0.43954 & 0.00147 &    5.08727 & 0.000015 & -0.155 &  2.88   \\
$f_{33}$ &  1.67322 & 0.00149 &   19.36597 & 0.000014 & -0.327 &  2.88   \\
$f_{34}$ &  0.25381 & 0.00151 &    2.93762 & 0.000015 & -0.472 &  2.90   \\
$f_{35}$ &  3.02746 & 0.00151 &   35.04005 & 0.000016 &  0.275 &  2.93   \\
$f_{36}$ &  1.91309 & 0.00159 &   22.14225 & 0.000012 & -0.022 &  2.83   \\
$f_{37}$ & 11.22560 & 0.00170 &  129.92593 & 0.000012 &  0.282 &  4.65   \\
$f_{38}$ &  2.89645 & 0.00173 &   33.52373 & 0.000013 & -0.272 &  2.63   \\
$f_{39}$ &  4.55862 & 0.00174 &   52.76181 & 0.000012 & -0.083 &  2.66   \\
$f_{40}$ &  4.76013 & 0.00174 &   55.09410 & 0.000012 &  0.346 &  3.10   \\
$f_{41}$ &  0.93383 & 0.00173 &   10.80822 & 0.000013 & -0.053 &  2.74   \\
$f_{42}$ &  1.42504 & 0.00173 &   16.49352 & 0.000013 &  0.278 &  2.77   \\
$f_{43}$ &  3.38899 & 0.00174 &   39.22442 & 0.000012 &  0.381 &  2.82   \\
$f_{44}$ &  0.14792 & 0.00177 &    1.71204 & 0.000012 & -0.493 &  2.79   \\
$f_{45}$ &  2.79809 & 0.00179 &   32.38530 & 0.000011 &  0.463 &  2.80   \\
$f_{46}$ &  0.61406 & 0.00183 &    7.10718 & 0.000013 &  0.236 &  2.78   \\
$f_{47}$ &  0.78215 & 0.00185 &    9.05266 & 0.000013 & -0.360 &  2.79   \\
$f_{48}$ &  1.76133 & 0.00189 &   20.38576 & 0.000011 &  0.025 &  2.79   \\
$f_{49}$ &  1.98728 & 0.00191 &   23.00093 & 0.000013 &  0.138 &  2.76   \\
$f_{50}$ &  1.58434 & 0.00195 &   18.33727 & 0.000011 &  0.015 &  2.77   \\
$f_{51}$ &  0.95913 & 0.00200 &   11.10104 & 0.000010 & -0.281 &  2.73   \\
$f_{52}$ &  4.43451 & 0.00202 &   51.32535 & 0.000010 & -0.318 &  2.83   \\
$f_{53}$ &  2.32122 & 0.00203 &   26.86597 & 0.000010 & -0.459 &  2.78   \\
$f_{54}$ &  1.13398 & 0.00204 &   13.12477 & 0.000010 &  0.437 &  2.80   \\
$f_{55}$ &  3.21544 & 0.00208 &   37.21574 & 0.000010 & -0.190 &  2.78   \\
$f_{56}$ &  8.77345 & 0.00217 &  101.54456 & 0.000009 &  0.458 &  3.38   \\
$f_{57}$ &  1.72569 & 0.00219 &   19.97326 & 0.000010 & -0.045 &  2.67   \\
$f_{58}$ &  1.85355 & 0.00213 &   21.45313 & 0.000010 & -0.492 &  2.80   \\
$f_{59}$ &  0.29858 & 0.00216 &    3.45579 & 0.000010 &  0.410 &  2.77   \\
\hline\\
$f_{60}$ &  1.02122 & 0.00218 &   11.81968 & 0.000009 & -0.250 &  2.80   \\
$f_{61}$ &  1.20689 & 0.00216 &   13.96863 & 0.000011 &  0.370 &  2.86   \\
$f_{62}$ &  3.91015 & 0.00219 &   45.25637 & 0.000009 &  0.381 &  2.74   \\
$f_{63}$ &  5.91544 & 0.00222 &   68.46574 & 0.000009 & -0.040 &  2.89   \\
$f_{64}$ &  1.24273 & 0.00226 &   14.38345 & 0.000010 & -0.160 &  2.85   \\
$f_{65}$ &  1.95592 & 0.00224 &   22.63796 & 0.000009 &  0.479 &  2.91   \\
$f_{66}$ &  1.31797 & 0.00228 &   15.25428 & 0.000009 & -0.195 &  2.91   \\
$f_{67}$ &  0.69020 & 0.00223 &    7.98843 & 0.000009 &  0.258 &  3.00   \\
\hline
\end{longtable}

\end{document}